\documentclass[bibyear,tradiabstract]{aa}
\usepackage{graphicx}
\usepackage[colorlinks=true,linkcolor=blue,allcolors=blue]{hyperref}%
\usepackage{natbib}
\usepackage{txfonts}
\begin{document} 

   \title{Evolution of gas envelopes and outgassed atmospheres \\of rocky planets formed via pebble accretion}

   \author{Piia Maria Tomberg
          \inst{1}
          \and
          Anders Johansen\inst{1}$^{,}$\inst{2}
          }

   \institute{Center for Star and Planet Formation, 
              Globe institute, University of Copenhagen,
              {\O}ster Voldgade 5-7, 1350  Copenhagen, Denmark\\
              \email{piiamaria.tomberg@sund.ku.dk}
        \and    
            Lund Observatory, Institute of Astronomy and 
            Theoretical Physics, 
            Lund University, Box 43, 221 00 Lund, Sweden
            }

   \date{Received 14.06.2024; accepted 11.09.2024}

 
  \abstract{
   We present here results of numerical simulations of the formation and early evolution of rocky planets through pebble accretion, with an with an emphasis on hydrogen envelope longevity and the composition of the outgassed atmosphere. We model planets with a range in mass from 0.1 to 5 Earth masses that orbit between 0.7 and 1.7 AU. The composition of the outgassed atmosphere is calculated with the partial pressure of free oxygen fit to geophysical models of magma ocean self-oxidation. XUV radiation powered photoevaporation is considered as the main driver of atmospheric escape. We model planets that remain below the pebble isolation mass and hence accrete tenuous envelopes only. We consider slow, medium or fast initial stellar rotation for the temporal evolution of the XUV flux. The loss of the envelope is a key event that allows the magma ocean to crystallise and outgas its bulk volatiles. The atmospheric composition of the majority of our simulated planets is dominated by CO$_2$. {Our planets} accrete a total of 11.6 Earth oceans of water, the majority of which enters the core. The hydrospheres of planets lighter than the Earth reach several times the mass of the Earth's modern oceans, while the hydrospheres of planets ranging from 1 to 3.5 Earth masses are comparable to those of our planet. However, planets of 4-5 Earth masses have smaller hydrospheres due to trapping of volatiles in their massive mantles. Overall, our simulations demonstrate that hydrogen envelopes are easily lost from rocky planets and that this envelope loss triggers the most primordial partitioning of volatiles between the solid mantle and the atmosphere.}


   \maketitle
%

\section{Introduction}
The \textit{James Webb} Space Telescope, with its marked increase in observing power compared to its predecessors, has the capability to characterise the atmospheres of small exoplanets \citep{Rotman+2023}. With the atmospheres of gas giant exoplanets already observed \citep{Constantinou+2023}, the first characterisation of a secondary atmosphere of high mean molecular weight could be imminent in the near future. Along with the recent detection of Earth-sized exoplanets such as the TRAPPIST-1 system \citep{Gillon+2017}, this prospect highlights the need for a comprehensive chemophysical atmospheric model linked to planet formation for planets in the habitable zone. The TRAPPIST-1b and TRAPPIST-1c planets have been shown to most likely lack atmospheres, despite their masses and effective temperatures being comparable to the rocky exoplanets of the Solar system \citep{Greene+2023,Zieba+2023}. The present lack of atmospheric observations for rocky planets imply that the terrestrial planets of the Solar System remain important calibration points for a complete model of early planetary evolution, including their early gas envelopes and atmospheres.\\
\\
The atmospheric evolution of rocky planets begins during their formation, as protoplanets grow by accreting pebbles \citep{Johansen+2015,Johansen+2021,Broz+2021,Olson+2021,Johansen+2023I} and by giant impacts \citep{Wetherill1990,Chambers&Wetherill1998,Woo+2022,Batygin&Morbidelli2023}. These planets consist of a mixture of metals and silicates surrounded by an envelope of hydrogen gas. In this work the words `envelope' and `atmosphere' are not used interchangeably, rather the envelope, consisting predominantly of hydrogen and helium \citep{Piso+Youdin2014}, sits on top of the atmosphere of heavier elements. As a growing planet is heated by the accretion heat, with some contribution also from short-lived radionuclides, the metals melt and sink to form a metallic core, while the silicates melt to form an overlying magma ocean \citep{Tronnes+2019}. The composition of the magma ocean defines the composition of the atmosphere underneath the hydrogen envelope, with the partial pressure of free oxygen over the magma ocean set by the contents of Fe$^{2+}$ and Fe$^{3+}$ in the magma ocean \citep{Armstrong+2019,Ortenzi+2020,Deng+2020}. As the magma ocean solidifies, it degasses much of its volatiles, which come to constitute the planet's most primordial atmosphere \citep{Elkins-Tanton2008,Deng+2020}. Such primordial atmospheres may later undergo significant changes due to plate tectonics \citep{Korenaga2013,Lammer+2018}, CO$_2$ cycling \citep{Berner2004,Tranvik+2009} and the emergence of life \citep{Lyons+2014,Krissansen-Totton+2018}.\\
\\
\noindent Both envelopes and atmospheres are nevertheless prone to mass loss after the protoplanetary disc phase. Atmospheric loss in planets is powered by a multitude of sources. The photoevaporation of the atmosphere due to the X-ray and UV radiation of the active young host star is believed to play an important role \citep{Owen&Wu2013,Erkaev+2014,Johnstone+2021}. Since H is the lightest element in the atmospheres, this element is most easily lost to photoevaporation, and by transfer of momentum to heavier gasses causes their loss as well \citep{Erkaev+2014}. Another physical mechanism is core powered mass loss, which heats the atmosphere due to the planet cooling from its initial accretional heat \citep{Gupta&Schlichting2019}. Small planets are normally not considered to maintain a hydrogen envelope \citep{Liggins+2022,Seligman+2022}, since such an envelope will be easily lost. One of the goals of this paper is to quantify the longevity of the hydrogen envelope of such small planets and how this depends on the planetary mass; this way we will also determine its effects on the crystallisation of the magma ocean and the outgassing of the atmosphere. 
\\
\\
The overarching aim of our paper is to show an in depth view into the early atmospheric outgassing and evolution of rocky planets orbiting a Sun-like star of varying stellar activity levels. This is achieved using the quantitative planet formation and evolution model created in \citet{Johansen+2023I}. We use the Accretion and Differentiation of Asteroids and Planets (ADAP) code detailed in that paper to simulate the evolution of terrestrial planets and the fates of their primary and secondary atmospheres at different orbital radii.\\
\\
The organisation of this paper is as follows. In section 2 we give insight to the parts of the ADAP code that most significantly impact a planet's primary and secondary atmospheres. In section 3 we show the outcomes of a range of simulations, focusing on the outcomes in regards to atmospheric composition and oxidation state. In section 4 we give our conclusions and discuss their implications for understanding the origin of planetary atmospheres.

\section{Methods}
\subsection{ADAP}
We simulate the formation of rocky planets and their atmospheres using the ADAP code developed for \cite{Johansen+2023I}. ADAP is a one-dimensional spherically symmetric code that numerically solves the accretion of material and the heat conduction through the interior layers for every time step. The code includes the melting of the planet and sinking of denser material as it melts, while conserving the total energy. It includes the heating from short-lived radiogenic heating, accretional heating and the heat released during differentiation.\\
\\
We use a volatile accretion scheme similar to that of \cite{Johansen+2023I} where water is accreted only up to the protoplanet mass of 0.01 M$_{\oplus}$. The envelope temperature at higher masses is greater than the ice sublimation temperature, which leads to loss of H$_2$O from the accreted pebbles to the flows that penetrate from the protoplanetary disc \citep{Johansen+2021,Wang+2023}. This approach yields a total water accretion of 11.6 Earth oceans independent of planet mass, with the majority of water residing in the core due to the siderophile nature of these volatiles \citep{Fischer+2020,Li+2020}. {We also model a set of completely dry planets to explore the results of a lack of water on the resulting atmospheres, in addition to a set of planets with 4 times the nominal water amount.} Our water accretion model provides a good description of Earth's and Mars' volatiles \citep{Johansen+2023III}. When we vary our planetary masses and orbital radii in this work, our results are therefore anchored in a relatively good match to the surface reservoir of an Earth-mass planet. We recognise that water delivery is in reality more complex and likely includes a contribution from icy asteroids from the outer regions of the asteroid belt \citep[e.g.][]{Raymond+2004}. The accretion of carbon and nitrogen also mimics the approach of \cite{Johansen+2023I} where these elements are delivered via organic molecules within the accreting pebbles that undergo sublimation and pyrolysis in the accretion process between 325 K and 425 K. Carbon is accreted until reaching a total mass of $4\times10^{-4}$\,M$_{\oplus}$ and nitrogen is accreted until its amount equals the Earth's atmospheric nitrogen reservoir of $4.02\times10^{18}\,$kg. This gives an atmospheric carbon budget that is comparable to that of the modern Venusian atmosphere. We compare here to Venus because Earth only carries an insignificant remnant of CO$_2$ in its atmosphere today, due to burial of this CO$_2$ by dissolution in the oceans and subduction into the mantle \citep{Dasgupta&Hirschmann2010}. {We do not vary the accretion budget of nitrogen or carbon here since while it would predictably affect eventual atmospheric composition, the majority of volatiles will be trapped in the metallic core, reducing the actual effect of a varied volatile budget.}
\begin{figure}
\centering
\includegraphics[width=9cm]{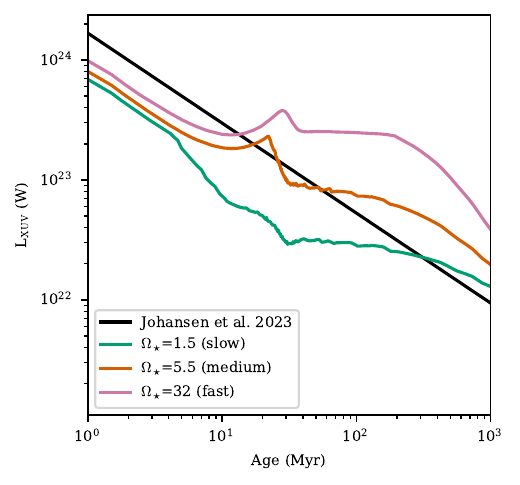}
   \caption{The XUV luminosity of \cite{Johansen+2023I} in black and of three stars with different initial rotation rates ($\Omega_{\odot}$) in units of the Sun's initial rotation rate ($\Omega_{\odot}$ = 2.67$\times10^{-6}$ rad/s) from \cite{Johnstone+2021}. These are slow, medium and fast rotators defined as the 5th 50th and 95th percentiles of their observed 150 Myr distribution.
           }
      \label{Fig:XUV}
\end{figure}
\begin{figure*}
\centering
\includegraphics[width=18cm]{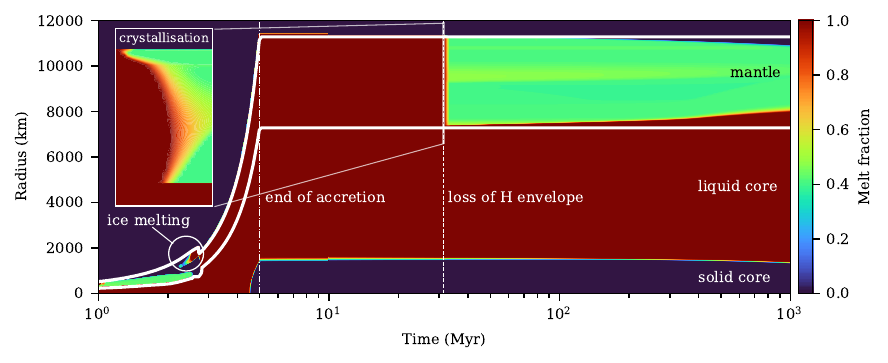}
   \caption{The interior melt fraction of a planet with mass $M$\,=\,4 M$_{\oplus}$ orbiting a medium rotator star at 1 AU. The colours show the melt fraction over time and radius with both the radius of the planet and the radius of the metallic core shown as a thick white line for reference. The end of accretion is marked with a vertical dash-dotted line at 5 Myr and the loss of the hydrogen envelope is marked with a vertical dashed line at 31.3 Myr. The magma ocean crystallises on a very short time-scale after 31.3 Myr as the hydrogen envelope is lost and its insulating effect disappears.{ The outgassing of volatiles, particularly CO$_2$ and N$_2$, takes place during the envelope's lifetime but the majority of the volatile budget, including H$_2$O, is rapidly degassed during the magma ocean solidification that takes place over 23 kyr here.} The base of the magma ocean nevertheless does not crystallise because it remains in contact with the core, which is stable to convection and hence loses its heat very slowly. The greatest amount of remnant accretion heat is stored in the outer core of the planet, since any accreted metals will sink through the liquid magma ocean and end up in the outer core. This results in a gradual reheating and melting of the lower mantle.
           }
      \label{Fig:meltfrac}
\end{figure*}
\subsection{Drivers of atmospheric escape}
In this work, all atmospheric molecules are assumed to be photolysed in the thermosphere, where from the escape of the resulting atoms is driven. {This simplification allows us to model atmospheric loss without the use of a photochemical model. In addition \cite{Johansen+2023III} showed that compared to keeping molecules intact, assuming atomisation only slightly increases the atmospheric mass loss rate.} The main method for the loss of the planets' envelope and atmosphere is photoevaporation due to the EUV and X-ray (collectively XUV) radiation from a young host star. {The mass loss efficiency coefficient in this work is $\eta=0.3$ \citep{Salz+2016}. Unlike some other papers such as \cite{Luger&Barnes2015} and \cite{Aguichine+2021} we set the effective radius at which the XUV photons are caught to a radius greater than the solid planet's radius, based on the parametrisation of \cite{Salz+2016}. This results in the more effective capture of XUV photons and higher levels of mass loss.} The strength of the XUV luminosity of a star depends on the initial rotation rate of the star. We take here a more realistic approach to the XUV mass loss than \cite{Johansen+2023I}, by adopting the evolution tracks given by \cite{Johnstone+2021} to determine the rate of atmospheric loss over time. The XUV luminosity used in the original ADAP calculations by \cite{Johansen+2023I} is shown in comparison to three stars of different rotation rates in Figure \ref{Fig:XUV}. \\
\\
The XUV luminosities of the three different stars were calculated using the Model For Rotation Of Stars (MORS) created by \cite{Johnstone+2021} and \cite{Spada+2013}. The MORS model calculates the time-dependent XUV luminosity of a star with a given mass, using the combined observations of young stellar clusters for short wavelengths (10-36 nm) and the Sun for longer wavelengths (36-92 nm). The relevant decay timescale of the  XUV luminosity depends on the initial rotation of the host star, as we illustrate in Figure \ref{Fig:XUV}.\\
\\
As a starting point for interpreting the planet formation simulations, we show in Figure \ref{Fig:meltfrac} the internal evolution of the melt fraction for a planet of mass $M=4\,\mathrm{M}_{\oplus}$ orbiting at 1 AU simulated using the ADAP code. We assume here, as in all our simulations, that the protoplanetary disc has a life-time of 5 million years and that the planetary accretion terminates after the dissipation of the protoplanetary disc. As is visible in Figure \ref{Fig:meltfrac}, the mantle of our super-Earth planet remains in the state of a magma ocean for the duration of the hydrogen envelope's lifetime. Thus, the hydrogen envelope must evaporate within the XUV-active timescale to facilitate the evolution of a secondary atmosphere. \\
\\
Another common mechanism of atmosphere and envelope loss is core-powered mass-loss, where atmospheric loss is driven by the gradual cooling of the planet \citep{Gupta&Schlichting2019}. In our work, however, we assume that the envelope loss is driven by XUV radiation and core-powered mass-loss is likely unimportant for the geometrically thin outgassed atmosphere with its high mean-molecular weight and low scale-height. In addition, our accreted envelope mass fractions are relatively low, further favouring photoevaporation relative to core-powered mass-loss \citep{Owen&Schlichting2024}. Thus, the effects of core-powered mass-loss are not considered in our simulations.\\
\\
Hydrogen in the outgassed atmosphere is exposed to XUV mass loss in the same way as hydrogen in the envelope, although the mean molecular weight of the atmosphere is much higher. The mechanism for atmospheric escape for the components heavier than hydrogen is through the drag of hydrogen atoms driving them upwards until eventual loss \citep{Zahnle&Kasting1986,Hunten+1987,Erkaev+2014,Tian2015}. Assuming a constant mixing ratio of H, C, N and O with height in the atmosphere, the escape rate of a species is calculated by checking whether the drag from the H atoms manages to accelerate them to their escape speed \citep{Johansen+2023II}. {The envelope, in contrast, is assumed to consist only of hydrogen as a simplification; including helium would not affect the envelope's thermal blanketing effect nor the ability of hydrogen to drive atmospheric loss.} 
\subsection{Outgassed atmospheres}
In the accretion phase, the planet differentiates very early as it is heated by short-lived radionuclides present in the silicates and by accretional heating. This is seen in Figure \ref{Fig:meltfrac}, where even during accretion there is a separate core and mantle, separated by the white line representing the core-mantle-boundary. This early differentiation is initially driven by the decay of $^{26}$Al. However, after a few million years this energy source is depleted and the mantle recrystallises. {The partition coefficient of a volatile between metal and silicates, $D_i = X_i^{\rm (met)}/X_i^{\rm (sil)}$ with $X_i$ denoting mass concentration of component $i$, depends on both temperature and pressure \citep{Grewal+2019,Li+2020,Fischer+2020}. We nevertheless fix the partition coefficient to $D_{\rm C}$=300, $D_{\rm N}$=10 and $D_{\rm H_2O}$=5. The uncertainty of these values was explored by varying them in \cite{Johansen+2023III}} A second differentiation stage starts as the melting of the water layer leads to the outgassing of a dense water atmosphere that insulates the planet to differentiate \citep{Johansen+2023II}. This phase is marked in Figure \ref{Fig:meltfrac} at approximately 3 Myr. Accretion terminates after 5 Myr when the protoplanetary disc is assumed to dissipate. However, this does not lead to immediate cooling of the magma ocean, since the hydrogen envelope acts as an insulator against heat loss. The mantle of the planet remains in a magma ocean state until the hydrogen envelope of the planet is lost due to XUV radiation at 31.3 Myr.
\\
\begin{figure}
\centering
\includegraphics[width=9cm]{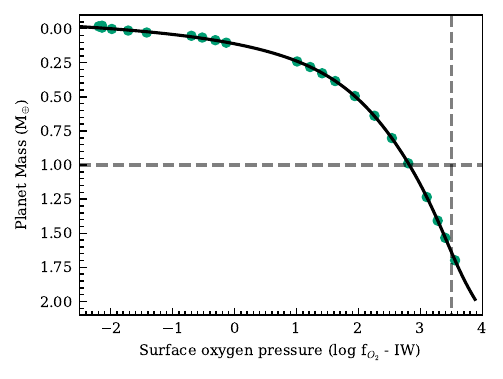}
   \caption{Our fit to the oxygen fugacity over the magma ocean as a function of planetary mass using a fifth-order polynomial. We use a surface temperature of 2100 K \citep{Deng+2020}. For reference, see Figure 3 of \citet{Deng+2020} where the oxygen fugacity is given as a function of the magma ocean base pressure instead. Earth's upper mantle surface redox state and Earth's mass are given in dashed lines. The polynomial constants are given in Table \ref{table:1}.}
      \label{Fig:IW_ME}
\end{figure}
\begin{figure*}
\centering
\includegraphics[width=18cm]{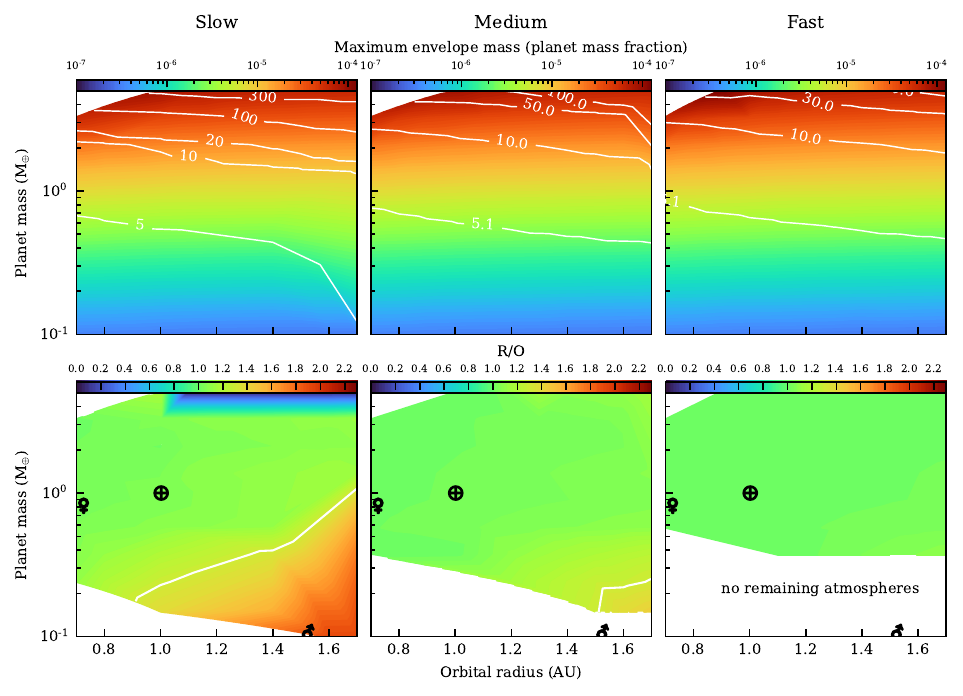}
   \caption{\textit{Upper panels:} The peak hydrogen envelope mass fraction at the end of accretion, before any mass loss, shown in colour for the slow, medium and fast host star rotators. The time when 90\% of the envelope mass is lost due to solar XUV radiation is shown in white contour lines in Myr. \textit{Lower panels:} The atomic composition of the outgassed atmospheres of the slow, medium and fast rotator simulations at 300 Myr. Values of R/O < 1 (blue) indicate an atmosphere with free oxygen, a value of R/O=1 (green) imply sufficient oxygen to oxidise H to H$_2$O and C to CO$_2$, and values of R/O > 1 (red) indicate reducing conditions {(see definition of R/O in equation \ref{eq:RO})}. The white contour lines in the lower panels show a value of R/O\,=\,1.3. The patch of R/O=0 present for the slow rotator is due to the perseverance of the envelope for those high planetary masses even after 300 Myr, with the magma ocean still providing a significant pressure of free oxygen.
           }
      \label{Fig:Menv_molecules}
\end{figure*}
\\
{The dissolution of volatiles in the magma ocean as it interacts with the atmosphere is set by the surface pressure. The surface pressure is too high to support atmospheric water while the envelope is still present and the magma ocean is open, therefore the water remains dissolved in the magma ocean. Since CO$_2$ has a low solubility in the magma ocean, it is mostly outgassed before the loss of the hydrogen envelope \citep{Johansen+2023III}. }\\
\\
{While the magma ocean is open, the outgassing of volatiles from the silicates is defined by the surface temperature and pressure as they relate to chemical equilibria.} In this work we use the thermochemical equilibrium speciation model of \cite{Ortenzi+2020} that describes the interaction between the open magma ocean and the atmosphere. They give the chemical equilibrium as the reaction set,
\begin{align}
    \mathrm{H}_2 + \frac{1}{2}\mathrm{O}_2 &\rightleftharpoons \mathrm{H}_2\mathrm{O} \label{eq1}\\
    \mathrm{CO} + \frac{1}{2}\mathrm{O}_2 &\rightleftharpoons \mathrm{CO}_2. \label{eq2}   
\end{align}
At lower temperatures of 873 -- 1873 K \citep{Tian&Heng2023} the following equilibriums become important as they transfer the C and N to CH$_4$ and NH$_3$ through the reactions,
\begin{align}
    \mathrm{CO} + 3 \mathrm{H}_2 &\rightleftharpoons \mathrm{CH}_4 + \mathrm{H}_2\mathrm{O} \label{eq3}\\
    \mathrm{N}_2 + 3\mathrm{H}_2 &\rightleftharpoons 2\mathrm{NH}_3. \label{eq4}
\end{align}
In oxidising conditions, with negligible free H$_2$, these reactions are not relevant. The CH$_4$ and NH$_3$ in our model would be susceptible to photolysis due to the high levels of XUV radiation of the first 300 Myr of the planets' lifetimes \citep{Kasting1982,Romanzin+2005,Kasting2014}. In addition, after envelope loss, the surface temperatures of our planets average 300-450\,K which makes the reformation of CH$_4$ unlikely as reactions (\ref{eq3}) and (\ref{eq4}) are kinetically inhibited at low temperatures. We therefore do not include equations (\ref{eq3}) and (\ref{eq4}) in our chemical equilibrium calculations, even under reducing conditions.
\begin{figure*}
\centering
\includegraphics[width=18cm]{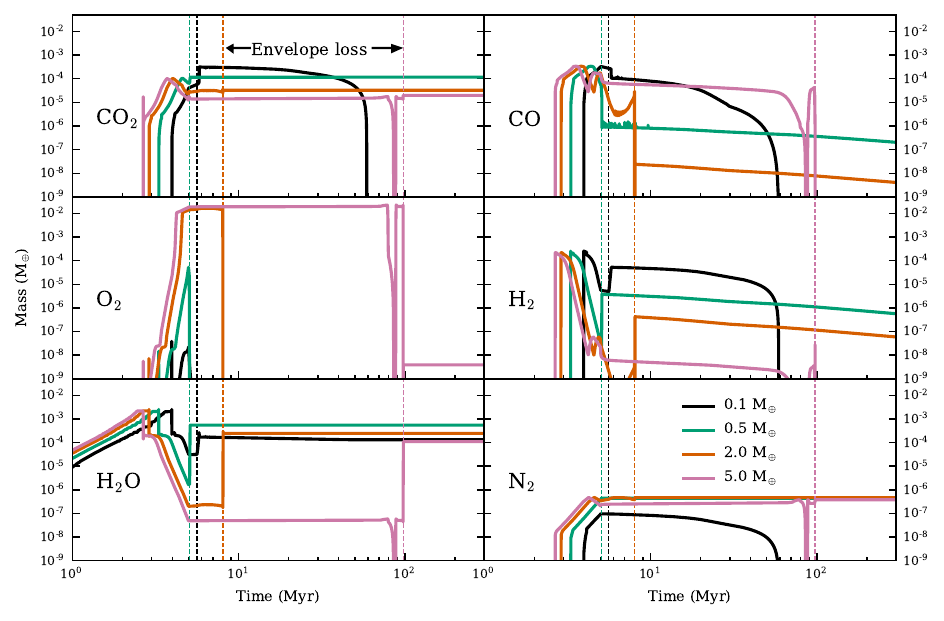}
   \caption{Evolution of the main degassed atmospheric components in four example planets of 0.1, 0.5, 2.0 and 5.0 Earth masses orbiting at 1 AU. The host star for these planets is the medium rotator. The water content includes both the atmospheric and surface water. The time of envelope loss is given in the respective colours of each planet mass in vertical dashed lines. The delay of features similar to those of low mass planets for more massive planets is due to the timing of the loss of the hydrogen envelope for these planets.}
      \label{Fig:atmcomp}
\end{figure*}
\\
\\
The oxidation state of molten mixtures of silicates and oxidised iron defines the partial pressure of oxygen in the melt (or, the oxygen fugacity, when referring to the activity rather than the pressure) and hence the pressure of free oxygen over the magma ocean \citep{Liggins+2022}. A common equilibrium state, or redox buffer, used as a scale in geochemistry is the iron-wüstite (IW) buffer, of which the Earth's upper mantle is 3-5 orders of magnitude more oxidising \citep{Hirschmann2022}. The oxygen fugacity describes the partial pressure of oxygen over the magma ocean and hence defines the speciation between H$_2$ and H$_2$O and between CO and CO$_2$ through equations (\ref{eq1}) and (\ref{eq2}), as discussed in \citet{Ortenzi+2020}. The main elements outgassed during the magma ocean stage are CO, H$_2$O and H$_2$ for a reduced magma and CO$_2$ and H$_2$O for an oxidised magma \citep{Ortenzi+2020}. The IW buffer is most relevant during the core formation stages of a magma ocean planet, when there is still free iron present in the silicates \citep{Huang+2021}, as it is represented by the equilibrium reaction
\begin{equation}\label{eq:freeiron}
    \mathrm{Fe} + \frac{1}{2} \mathrm{O}_2 \rightleftharpoons \mathrm{FeO}.
\end{equation}
\citet{Armstrong+2019} showed that the partial pressure of free O$_2$ in the magma ocean increases strongly with increasing pressure within the magma. After the end of accretion, the magma ocean is quickly emptied of metallic iron. The free O$_2$ in equation (\ref{eq:freeiron}) then feeds into the reaction for the further oxidation of iron,
\begin{equation}\label{eq:nofreeiron}
    \mathrm{FeO} + \frac{1}{4}\mathrm{O}_2 \rightleftharpoons \mathrm{FeO}_{1.5}.
\end{equation}
Here, FeO$_{1.5}$ represents Fe$^{3+}$. The fraction of Fe$^{3+}$/Fe$^{2+}$ is set at the core-mantle boundary and kept constant throughout the magma ocean by convection. This yields a strongly oxidised upper magma ocean. The oxidation state of the upper magma ocean, in the absence of metallic iron, is thus set by the oxygen fugacity
\begin{equation}
    f_{\mathrm{O}_2}=\left(\frac{a^{\mathrm{melt}}_{\mathrm{FeO_{1.5}}}}
    {a^{\mathrm{melt}}_{\mathrm{FeO}}\times K}\right)^4,
\end{equation}
where $K$ is the equilibrium constant of equation (\ref{eq:nofreeiron}) and $a_{\mathrm{FeO}}$ and $a_{\mathrm{FeO}_{1.5}}$ are the concentrations of the relevant molecules in the silicate melt. \citet{Deng+2020} used calculations of the two oxygen reactions to tie the oxygen fugacity of a fully molten silicate mantle to the pressure at the bottom of the magma ocean for objects ranging from Earth to Moon masses. In Figure \ref{Fig:IW_ME} we relate the magma ocean base pressure to the planetary mass with a fifth-order polynomial to the results of \cite{Deng+2020} to obtain an oxygen fugacity that is dependent on the planet's mass. Any planets of a greater mass than 2 M$_{\oplus}$ were set to have a constant value of $\Delta$IW=+4, in relation to the IW buffer, due to the mass-balance limitations of the Deng model and to prevent unphysical runaway oxygen outgassing. The atmospheric C and H is anyway fully in the form of CO$_2$ and H$_2$O  for such high oxygen fugacities. The SiO pressure over the magma is given by a saturated vapour pressure approach \citep{Johansen+2023II}.\\
\begin{table}[]
\centering
\caption{Constants used in the fifth-order polynomial fit to the oxygen fugacity model of \cite{Deng+2020}.}
\begin{tabular}{l|l}\label{table:1}
 & Constant \\ \hline
C$_1$ & 2.55893462$\times10^{-9}$  \\ \hline
C$_2$ & -8.26050794$\times10^{-7}$ \\ \hline
C$_3$ & 1.07405538$\times10^{-4}$\\ \hline
C$_4$ & -7.22411940$\times10^{-3}$\\ \hline
C$_5$ & 2.87447212$\times10^{-1}$\\ \hline
C$_6$ & 3.01444278
\end{tabular}
\end{table}
\\
{Immediately after envelope loss the magma ocean rapidly solidifies and} degasses both water and carbon-containing molecules \citep{Lichtenberg+2021}. Due to the high temperatures reached in the magma ocean, water is initially degassed in vapour form and later mostly condenses onto the surface. Under reducing conditions, hydrogen is also degassed \citep{Ortenzi+2020} but due to its low mass H is susceptible to atmospheric loss by the impending XUV radiation of the host star. Water is both degassed during magma ocean solidification and retained in the mantle as OH compounds, leaving a significant amount of water trapped in the mantle of the planet. The minerals that accept OH are saturated at different limiting values and the water saturation of silicates is therefore strongly pressure dependent. Since carbon is much less soluble in the mantle, it is degassed almost in its entirety. We use here the approach of \cite{Elkins-Tanton2008} to calculate the partitioning of volatiles between melt and solid.
\subsection{Simulation setup}
Our simulations consider planets ranging from 0.1 to 5 Earth masses, bounded at the upper end by the pebble isolation mass, which is 4.62 Earth masses at 1 AU \citep{Bitsch+2018}. These planets therefore accrete only tenuous envelopes, since efficient cooling and envelope contraction is prevented by the continuous release of accretion energy at the surface. {This additional heating from infalling pebbles is the leading cause of the low initial envelope masses in this work, in contrast to \cite{Lambrechts+2019}, \cite{Mordasini+2020} and \cite{Ormel+2021} who do not take pebble accretion into account.} Most of our simulations are run with conditions that include water, but we run a small subset without considering water to mimic conditions interior of the water ice line. The host star's parameters are varied by the three XUV activity levels given in Figure \ref{Fig:XUV}, and the orbital radius of the planets is varied from 0.7 to 1.7 AU. The ice line was likely situated slightly interior of 1 AU during most of the evolution of the protoplanetary disc \citep{Mori+2021}. On the other hand, under very cold conditions the water ice line in the envelope of the planet moves from the recycling zone down to the bound envelope \citep{Wang+2023}. Then our assumption that water is lost during the accretion is no longer valid \citep{Johansen+2021}. Hence, we do not consider planets that accrete significant enough amounts of ice to grow to water worlds or ice giants.

\section{Results}
\subsection{Loss of hydrogen envelope}
A clear result of our work is that for all planets the envelope is lost roughly within 300 million years, although the higher mass planets orbiting a slow rotator star need more time to achieve this. The loss of the envelope is presented in Figure \ref{Fig:Menv_molecules}. The upper panels of Figure \ref{Fig:Menv_molecules} show the hydrogen envelopes' maximum accreted mass fractions and life-times in white contour lines. The rotator speed of the host star is a crucial factor in envelope l{o}ss timescale, where the high mass planets orbiting the most active star take at most 65\,Myr to lose their envelopes, while planets of $M=5$\,M$_{\oplus}$ orbiting a slow rotating star need more than 300\,Myr to do so.
\begin{figure}
\centering
\includegraphics[width=9cm]{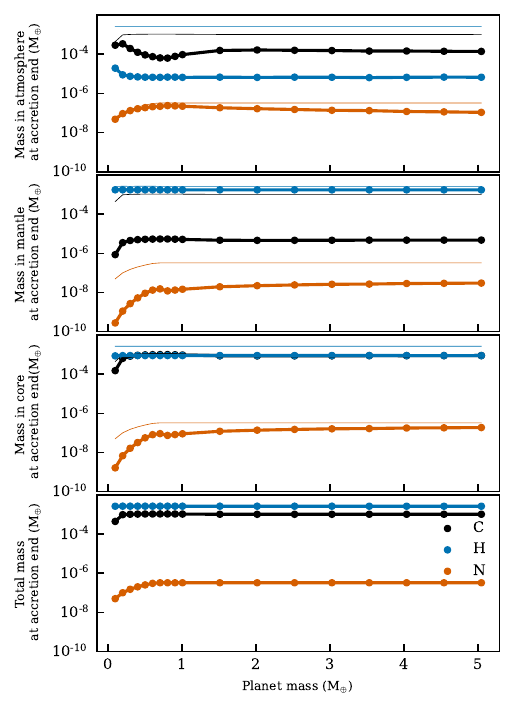}
   \caption{The masses of hydrogen, carbon and nitrogen at the end of accretion (but before magma ocean crystallisation) in the atmospheres (top), mantles and cores (middle) and total (bottom) of planets at 1 AU, as a function of the planetary mass, as set by pebble accretion. {Thin lines in the top three panels indicate the total budget for reference.} We see here that the greatest variation with planet mass occurs in the low mass planets, while that variation is still low. {Planets of greater than 1 Earth mass have the same initial volatile masses since these are bounded by the high temperatures of the envelope limiting volatile accretion, as described in section 2.1.} Therefore any differences in atmospheric composition by planet mass must be the result of processes taking place after accretion ends, as volatile accretion is relatively similar for all planet masses, which is a consequence of our envelope processing approach to volatile accretion \citep{Johansen+2023III}.
           }
      \label{Fig:HCN}
\end{figure}
\begin{figure*}
\centering
\includegraphics[width=18cm]{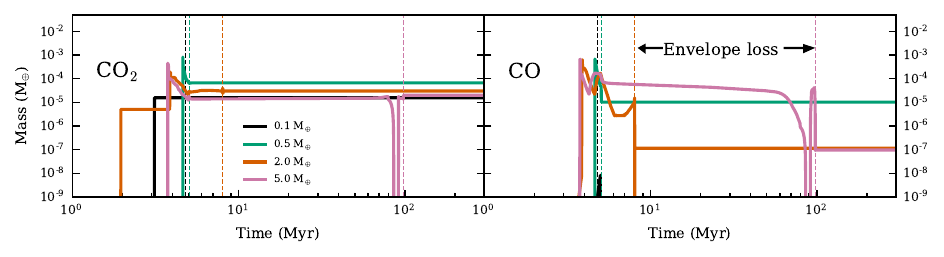}
   \caption{Evolution of the main degassed atmospheric components in four example planets formed with dry accretion. The planet with M = 2 M$_{\oplus}$ in this {dry} case has {more} CO than its wet accretion counterpart because the dry planet is {hotter and under the hot conditions CO$_2$ tends to shift more to CO}. The similarities to the wet accretion case imply that completely dry accretion does not greatly affect the atmospheric composition of the planets.}
      \label{Fig:atmcomp_dry}
\end{figure*}
\begin{figure*}
\centering
\includegraphics[width=18cm]{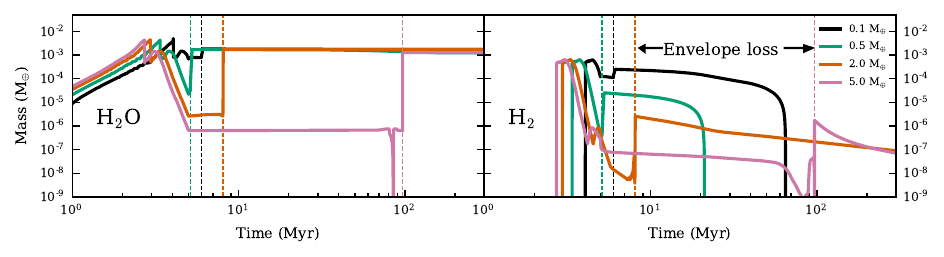}
   \caption{{Evolution of the degassed atmospheric components in four example planets formed with 4 times the initial water budget. The main difference from the nominal simulations in Figure \ref{Fig:atmcomp} is the increased H$_2$O and H$_2$ content in the atmosphere after envelope loss, other atmospheric components remain unchanged. In addition, in this increased water accretion case, the 0.5 Earth mass planet also loses its entire atmosphere due to the increased presence of H$_2$ driving atmospheric loss more easily.
   }}
      \label{Fig:atmcomp_4xh2o}
\end{figure*}
\subsection{Effects of the hydrogen envelope}
As is apparent from Figure \ref{Fig:meltfrac}, which exemplifies the evolution of a planet with $M$\,=\,4 M$_{\oplus}$, the planet maintains a liquid magma ocean until the envelope is lost. The magma ocean solidifies quickly after envelope loss and only then outgasses a significant atmosphere. The lower mantle nevertheless remelts slowly due to the heating from the outer core where the majority of accretional heat is stored. This shows how the presence of a hydrogen envelope dictates the surface conditions of a planet. The evolution of a degassed atmosphere is put to a pause until the complete loss of the hydrogen envelope, which, depending on the host star rotation speeds, can take more than 300\,Myr. 

\subsection{Outgassed atmosphere composition}
The lower panels of Figure \ref{Fig:Menv_molecules} show the final atmospheric composition of the planets. We define here a fictitious atomic ratio,
\begin{equation}\label{eq:RO}
    \mathrm{R}/\mathrm{O}=\frac{2\mathrm{C}+\frac{1}{2}\mathrm{H}}{\mathrm{O}}
\end{equation}
where values of R/O < 1 indicate an atmosphere with free oxygen, a values R/O = 1 shows sufficient oxygen to oxidise H to H$_2$O and C to CO$_2$, and values R/O > 1 indicate a reducing atmosphere \citep{Johansen+2024}. There, we find that with decreasing host star activity, low-mass planets at large orbital radii can maintain reducing atmospheres. Also, with increasing host star activity, more of the low mass planets lose their atmospheres entirely, as shown by the white areas in the figure. In Figure \ref{Fig:atmcomp}, the evolution of the atmospheric compositions of several representative planets orbiting a medium rotating star at 1 AU is shown.\\
\\
The hydrogen envelope exerts a remarkable amount of pressure on the early outgassed atmosphere that is rich in O$_2$, CO$_2$ and initially also SiO. As the early atmosphere cools and the SiO is incorporated into the magma ocean, the atmosphere is compressed and CO$_2$ begins to experience reabsorption into the magma ocean when exceeding the equilibrium condition
\begin{equation}
    X_{\mathrm{CO}_2}<
    4.4\times10^{12}P_{\mathrm{CO}_2}(\mathrm{Pa}),
\end{equation}
where $X_{\mathrm{CO}_2}$ is the mass fraction of CO$_2$ in the magma ocean and $P_{\mathrm{CO}_2}$ is the pressure exerted on the outgassed CO$_2$ \citep{Hirschmann+2012}. A similar effect happens for O$_2$ and H$_2$O. This effect of early atmosphere reabsorption is visible for the planet with $M$\,=\,5 M$_{\oplus}$ in Figure \ref{Fig:atmcomp}, where the mass of all atmospheric elements fall temporarily to zero before eventual envelope loss.
\\
\\
Since the atmospheric composition changes very significantly with planet mass, while it depends only very little on the distance from the star, an in-depth analysis on atmospheric composition is only performed on planets orbiting a star of medium XUV activity levels at 1\,AU. From Figure \ref{Fig:atmcomp} it is apparent that the lowest mass planets of $M$\,=\,0.1 M$_{\oplus}$ lose their atmospheres entirely. However, the atmospheres of higher mass planets are massive enough to remain despite the XUV irradiation. Another notable difference for the lightest planets is the increased initial CO$_2$ in the atmosphere {as seen in the early atmospheric CO$_2$ panel of Figure \ref{Fig:atmcomp}}. This is caused by the smaller core that is not able to absorb as much CO$_2$ from the mantle as their more massive counterparts, leaving more of it to be later degassed as the magma ocean solidifies.\\
\\
The mass budgets in the atmosphere, mantle and core of C, H and N at the end of accretion for the simulated planet mass range is given in Figure \ref{Fig:HCN}. The initially accreted mass of hydrogen varies little, while carbon and nitrogen see the most variation for low mass planets $<1$M$_{\oplus}$. This is due to the higher vaporisation temperature of C- and N-bearing organics and graphite, which implies that the C and N budgets continue to increase up to a planetary mass of 0.1 M$_{\oplus}$.\\
\\
The amount of available oxygen sets the scene for the remaining atmospheric composition of the different mass planets through oxidation. The primary source for the oxygen is the self-oxidation of the magma ocean which leads to the outgassing of significant O$_2$ from the magma ocean. This oxygen comes from the accreted FeO, which may in turn have formed in planetesimals and protoplanets by reaction of metallic iron with water H$_2$O \citep{Johansen+2023I}. This leads to the majority of simulated planets having an atmosphere that favours CO$_2$ over CO and H$_2$O over H$_2$. As is visible in the middle-left panel of Figure \ref{Fig:atmcomp} the $M$\,=\,5\,M$_{\oplus}$ mass planet ends up with a small remnant of free oxygen due to the photodissociation of H$_2$O, which then oxidises the atmosphere fully. For the medium mass planets the atmosphere contains enough oxygen to have significant CO$_2$, while CO is also present, with a similar trend for H$_2$O and H$_2$. \\
\\
The amount of nitrogen in the atmospheres of planets remains relatively independent of whether the planet is in a magma ocean stage or a solid mantle stage, because of the low solubility of nitrogen in the magma \citep{Sossi+2020}. Since N does not interact with oxygen, carbon or hydrogen in our chemical equilibrium model, it stays as a neutral part of the early atmosphere. The amount of hydrogen that remains in the atmospheres of the planets is negligible as it is slowly but steadily lost due to XUV radiation after envelope loss.

\subsection{Dry accretion}
In order to probe the atmospheric composition of planets that form interior of the water ice line, a set of planets orbiting at 1 AU around a medium rotating star were simulated with completely dry accretion. We maintain the same accreted amount of C and N, since their refractory organics hosts survive well interior of the water ice line \citep{Johansen+2023III}. In Figure \ref{Fig:atmcomp_dry} the atmospheric evolution of these planets is given, showing only the most relevant molecules: CO and CO$_2$. The timing of envelope loss is not affected by the change in accretion type. \\
\\
A notable difference from the wet accretion planets is a complete suppression of atmospheric loss for all planets. Since there is no atmospheric H$_2$ or H$_2$O, H cannot act as a proxy for energy transfer for the heavier molecules, thus preventing any atmospheric loss for planets that lose their atmospheres with wet accretion. There is no H$_2$ in the atmospheres here due to the lack of water which would otherwise produce H$_2$ by outgassing from the magma ocean under reducing conditions.\\
\\
The interiors of the dry planets differ very little from the wet planets. The atmospheres of the dry planets lack water and therefore the surface temperature of these planets is higher since the lack of moist convection leads to hotter conditions with its adiabatic atmosphere structure. {Despite the lack of a water vapour induced greenhouse effect, the dry planet therefore reaches a higher surface temperature from the CO$_2$ greenhouse effect alone. Note that we kept the oxidation state of the mantle unchanged in the dry models, despite the possibility that the source of FeO in the magma ocean was early-accreted water \citep{Johansen+2023I}.} The cores and mantles of the dry planets are the same size as their wet counterparts.
{
\subsection{Enhanced water accretion}
In order to explore the effects that an increased water accretion regime would have on the resulting atmospheres, we simulated a set of planets that accrete 4 times the nominal amount of water. This leads to the accretion of 0.04 Earth masses of water for all planets in the mass range of 0.1 to 5 Earth masses.\\
\\
An expected difference from the nominal simulations in Figure \ref{Fig:atmcomp} is the increased H$_2$O and H$_2$ content in this increased water accretion case. The increased water budget leads to more water dissociation, giving an atmosphere more enriched in H$_2$. A notable difference from the other water accretion cases is a complete loss of atmosphere for not only the 0.1 Earth mass planet, but also the 0.5 Earth mass planet. This is due to the increased presence of hydrogen leading to the more efficient excitation of the other atmospheric components, leading to atmospheric loss due to XUV irradiation.
}
\begin{figure}
\centering
\includegraphics[width=9cm]{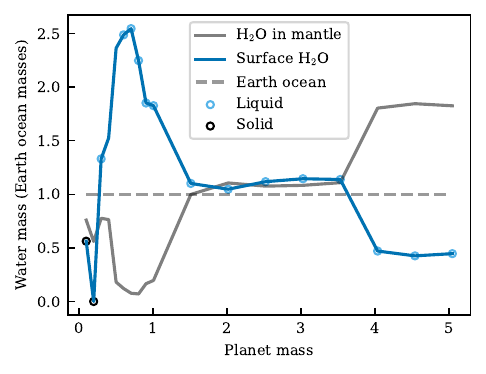}
   \caption{Masses of the hydrospheres of all simulations performed at 1 AU. The colour of the circles indicate whether the surface conditions of the planet at 300 Myr allow for water to be in a solid or liquid form with black indicating solid and light blue liquid. The mass of the Earth's oceans is given with a horisontal dashed line.}
      \label{Fig:hydro}
\end{figure} 
\begin{figure*}
\centering
\includegraphics[width=18cm]{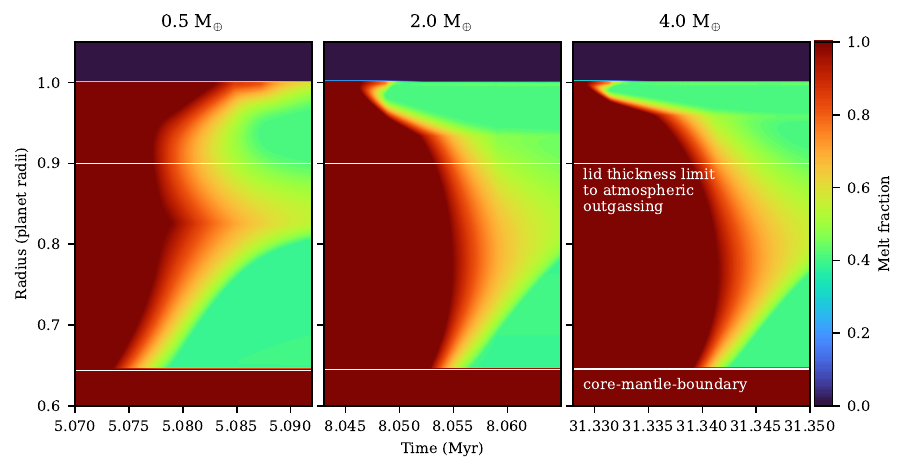}
   \caption{The melt fraction of different mass planets at 1\,AU during crystallisation, showing the variance in crystallisation lid regimes. The upper white horisontal line represents the lid thickness at which degassing is no longer efficient and volatiles start to be trapped in the mantle. The lower white line shows the core-mantle-boundary. The $M$\,=\,0.5\,M$_{\oplus}$ planet has a short magma ocean cooling time-scale, resulting in bottom-up crystallisation without forming a lid. The other two planets first form a thin lid before also starting to solidify from the bottom. The planet with $M$\,=\,2\,M$_{\oplus}$ planet solidifies slightly faster than the $M$\,=\,4\,M$_{\oplus}$ planet, which allows for more of the volatiles to be degassed through the thin lid before the lid grows thicker than our assumed limit for where degassing can occur.}
      \label{Fig:crystallisation}
\end{figure*}
\begin{figure}
\centering
\includegraphics[width=9cm]{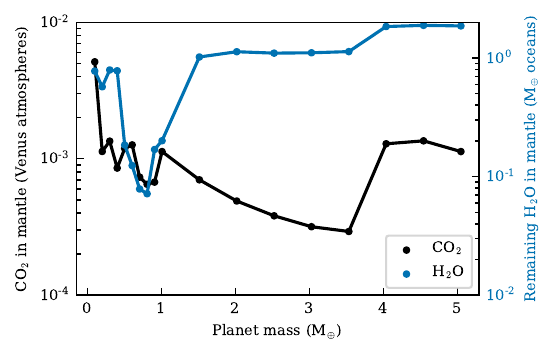}
   \caption{The mass of CO$_2$ and H$_2$O remaining in the mantles of different mass planets after the solidification of the magma ocean. It is evident that low mass planets tend to outgas water much more efficiently than high mass planets. While the variation in outgassed CO$_2$ is not as large as for H$_2$O, due to the lower solubility of CO$_2$ in magma, the tendency for high-mass planets to retain a larger mantle volatile reservoir extends to CO$_2$ as well.}
      \label{Fig:reservoir}
\end{figure}
\subsection{Hydrosphere}
Figure \ref{Fig:hydro} shows the masses of the surface and mantle water reservoirs of planets with varying masses. The planets with lowest masses end up with very little surface water due to the outgassing of H$_2$ and subsequent atmospheric loss, but the planets ranging from 0.5 to 1.0 M$_{\oplus}$ maintain significant hydrospheres. Planets with masses between 1.5 and 3.5 Earth masses maintain hydrospheres similar to the Earth's, while the highest mass planets show a deficit of surface water relative to their lower-mass counterparts. \\
\\
The thickness of a silicate lid on top of the magma ocean is one of the main limiting factors of volatile degassing. We consider a lid with the thickness of 10\% of the planet's radius to be limiting to the point of not allowing outgassing{, despite \cite{Dorn+2018} showing that the choice of lid thickness does not have as much of an effect on outgassing as planet mass does}. The retention of water by high-mass planets is due to the formation of a lid on the magma ocean before the crystallisation of the lower layers, as seen in Figure \ref{Fig:crystallisation}. Planets of low mass crystallise their magma ocean from the bottom up due to the short cooling time-scales of their smaller mantles. Planets of masses $M$\,$\geq 4$M$_{\oplus}$ do not experience such an early magma ocean base crystallisation and leave a significant amount of water frozen into the mantle.\\   
\\
Another factor in the eventual surface water content is the solubility of water in the magma ocean. Both CO$_2$ and H$_2$O are degassed during magma ocean solidification, but the former is degassed faster due to its lower solubility in silicates. The solidification of the magma ocean releases varying amounts of volatiles, depending on the thickness of the mantle. In Figure \ref{Fig:reservoir} the remaining CO$_2$ and H$_2$O after solidification in the mantle are shown for various planet masses. This shows how planets more massive than 4 Earth masses have significant amounts of water frozen into their mantles, while CO$_2$ is more readily degassed. These trapped volatiles are likely later outgassed by volcanism once the majority of the mantle has remelted and plate tectonics begin to occur.

\section{Discussion and conclusions}

\noindent 
The simulation code that we use in this paper must necessarily choose specific model parameters for planetary accretion, differentiation, outgassing, planetary chemistry as well as other processes. We base our calculations here on the model of \cite{Johansen+2023I} with some additional modifications made for some aspects such as a more realistic approach to XUV mass loss. The exclusion of alternative interior and atmosphere models from this work introduces an uncertainty that cannot be easily quantified. However, we believe that the choices of physical and chemical parameters made here are well-founded and thus give a good insight into the accretion and early evolution of rocky planets and their atmospheres. Particularly, the variations in stellar rotation and planetary mass and orbit give a rough estimate for the effect that parameter choices such as planetary mass and mass-loss rates have on the results.\\
\\
A key result of this paper is that the warming effect of the early hydrogen envelope keeps the surface of a planet in the phase of a magma ocean until the envelope is lost. This implies that the timing of primary atmosphere loss {due to stellar XUV radiation} can introduce a significant delay in the evolution of rocky planets. This delay is seen most strongly for massive planets at large orbital distances (see Figure \ref{Fig:Menv_molecules}) and can be up to 60 Myr for fast rotating host stars, 120 Myr for medium rotators and greater than 300 Myr for planets orbiting a slow rotator star. These conclusions are similar to those of \cite{Kite+Barnett2020} who also found that hot super-Earths tend to form with thick hydrogen envelopes that enforce a state of a molten magma ocean mantle until the loss of these envelopes. {During this magma ocean stage only carbon-containing volatiles are outgassed, while the water remains trapped in the molten magma.}\\
\\
As the magma ocean {rapidly} solidifies, the magma degasses H$_2$, H$_2$O, CO, CO$_2$ and N$_2$. Low mass planets (with masses up to 0.7 Earth masses) around a slow-rotating host star that do not lose their atmospheres to the XUV-powered photoevaporation retain reducing atmospheres with significant amounts of H$_2$. However,  planets around medium and fast rotating stars, as well as intermediate and high mass planets around slowly rotating stars, all end up with oxidising atmospheres. \\
\\
We performed an additional series of simulations of planets that form dry, to mimic accretion conditions interior of the water ice line. The timing of envelope loss as well as the atmospheric compositions in terms of CO and CO$_2$ are very similar. The most significant difference between dry and wet accretion is that the lack of H$_2$ in the dry planets does not allow for any atmospheric loss, because of the absence of H$_2$ that would have otherwise lifted the heavier molecules to escape from the planet. We also performed a set of simulations for planets that accrete 4 times the nominal amount of water, which end up with higher water and therefore hydrogen content, which drive more intensive atmospheric loss than the nominal case.\\
\\
The hydrospheres of mid-mass planets (those ranging in mass from 1 to 3.5 Earth masses) are similar in mass to the Earth's oceans, while high-mass planets are more deficient in water. This is due to more of the water getting trapped in the massive magma oceans during solidification. Low-mass planets that retain their atmospheres, however, have higher amounts of surface water compared to the Earth. This could indicate an increased likelihood of low-mass rocky planets in the habitable zone to form with surface water budgets significantly above the terrestrial value. Our findings that planets in the habitable zone may become endowed with a wide range of surface water even from the magma ocean outgassing alone thus have implications both for the possible presence of water worlds orbiting other stars and for the surface conditions during the origin of life.

\begin{acknowledgements}
    A.J. acknowledges funding from the Danish National Research Foundation (DNRF Chair Grant DNRF159), the Carlsberg Foundation (Semper Ardens: Advance grant FIRSTATMO), the Knut and Alice Wallenberg  Foundation (Wallenberg Scholar Grant 2019.0442) and the Göran Gustafsson Foundation. The computations were enabled by resources provided by the Swedish National Infrastructure for Computing (SNIC), partially funded by the Swedish Research Council through grant agreement no. 2020/5-387. 
\end{acknowledgements}

\end{document}